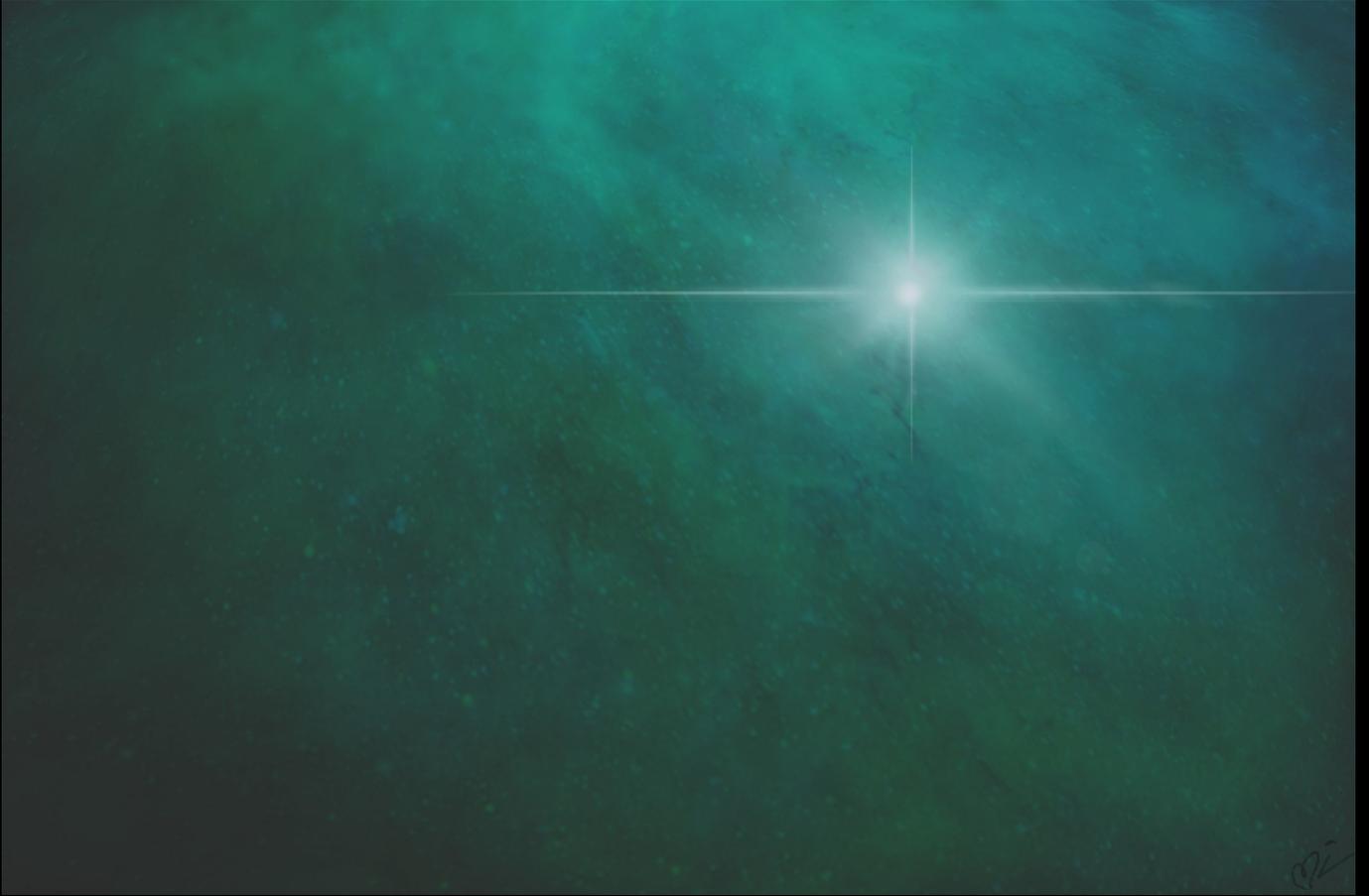

# Proceedings of the ETHICOMP 2022

Effectiveness of ICT ethics – How do we help solve ethical problems in the field of ICT?

Eds. Jani Koskinen, Kai K. Kimppa, Olli Heimo, Juhani Naskali, Salla Ponkala and Minna M. Rantanen
UNIVERSITY OF TURKU, Turku, Finland

# Proceedings of the ETHICOMP 2022

## Effectiveness of ICT ethics – How do we help solve ethical problems in the field of ICT?

**Editors:**

Jani Koskinen, Kai K. Kimppa, Olli Heimo, Juhani Naskali, Salla Ponkala and Minna M. Rantanen

**Publisher:**

University of Turku, Turku, Finland

**Copyrights:**

**Cover art**

"Northern Star" by Minna M. Rantanen

**ISBN** 978-951-29-8989-8  (PDF)

# Building the Learning Environment for Sustainable Development: a Co-creation approach

Ewa Duda[0000-0003-4535-6388]

Maria Grzegorzewska University, Warsaw, Poland

eduda@aps.edu.pl

**Abstract.** Education for sustainable development supports the improvement of knowledge, skills, attitudes and behaviors related to global challenges such as climate change, global warming and environmental degradation, among others. It is increasingly taking place through projects based on information and communication technologies. The effectiveness of the actions taken depends not only on the quality of the project activities or the sophistication of the innovative tools used. Social commitment also depends on the beliefs and moral judgements manifested by potential recipients of educational activities on environmental issues. This study aimed to identify the beliefs and moral judgements that may facilitate or hinder the implementation of educational activities based on information and communication technology, shaping pro-environmental attitudes and behavior among city dwellers. Based on the co-creation workshops conducted, five general categories emerged: responsibility, sense of empowerment, local leadership, real eco-approach, and eco-knowledge. The research findings may contribute to the design of educational activities dedicated to shaping the pro-environmental behavior of city dwellers.

## 1    Introduction

One of the biggest challenges of recent years has become the growing need to slow down the direction of human-caused climate change. Issues such as environmental degradation, negative consequences of globalization, social inequality and poverty are increasingly noticeable. We are beginning to understand we should be more concerned about the quality of the environment, but above all, be aware of how not to destroy what nature offers us and give the next generation a chance to (worthily) live. Transformations, both systemic and individual, are necessary. Awareness of the negative impacts of our acts plays a significant role in changing current unsuitable habits. It is essential to learn how to anticipate the consequences of future actions and start planning them more carefully.

One of the tools to support human behavior change is the creation of proenvironmental policies based on education. In particular, education for sustainable development (ESD) (Vare and Scott, 2007; Leicht, Heiss and Byun, 2018; Scott and Vare, 2020) is becoming more widely recognized and implemented in everyday life. ESD is much more than environmental education. It has a broader, multidisciplinary context, including sociological, pedagogical, economic, political, and cultural aspects (Arbuthnott, 2009; Venkataraman, 2010). Education for sustainable development refers not only to environmental knowledge but also to relevant skills, attitudes, and beliefs. Values such as sustainable society, responsibility for the Earth, and responsibility for a shared future are becoming increasingly important. However, it should be noted that they are still not sufficiently reflected in our daily routine. A solid education on a global scale seems to be necessary.

ESD is constantly evolving, covering more and more issues and dimensions. The approach applies at different levels, from early childhood education (Hedefalk et al., 2015; Siraj-Blatchford et al., 2016), school education (Mogren and Gericke, 2017; Hallinger and Nguyen, 2020), higher education (Mulà et al., 2017; Hallinger and Chatpinyakoop, 2019), but also within adult learning (Noguchi, 2019). The potential of this last target group seems underestimated and not fully explored. Moreover, the main focus of ESD is on formal education, leaving behind the non-formal and informal learning that fills most of our lives. Without any doubt, more attention should be paid to adult education for sustainable development.

Educational pro-environmental activities based on information and communication technology (ICT) are one example of the ESD concept implementation. Projects aimed at enhancing people's environment-friendly behavior attempt to involve adults in the learning process in innovative ways. Such projects adopt the principles of encouragement through a system of incentives rather than penalties. The growing number of community-based programs not only enables learners to develop their everyday educational practices but also provides an opportunity to deepen research in the field of a learning environment. Researchers are expanding their analysis by addressing issues of pro-environmental behavior, sustainability interventions (Ro, 2017), social learning (Kaaronen & Strelkovskii, 2020), and environmental knowledge (Pothitou et al., 2016). The conducted research allows gaining knowledge about people's habits, attitudes, perceptions, and values regarding ecology issues.

The presented article reflects on research to construct the scientific background for an emerging transdisciplinary project designing an interactive ICT system to promote pro-environmental behavior among urban residents. It focuses on identifying the existing beliefs and moral judgements of people involved in a participatory learning process based on a co-creation approach. Their recognition and understanding will contribute to a better implementation of the process of co-creating an effective lifelong learning environment for climate improvement. The consecutive paragraph will briefly present the theoretical background and literature review on moral judgements. The third paragraph presents the study methodology, the fourth one shows the results obtained during the co-creation workshop, and the fifth one discusses the findings and conclusions. The article closes with references.

## 2      Theoretical background

### 2.1    Co-creation

Many researchers and practitioners consider the co-creation method as a valuable tool for involving future audiences in the process of designing products (Roberts & Darler, 2017), services (Jaakkola et al., 2015), and values (Farr, 2015). In particular, the potential of using co-creation is widely considered in building ICT-based solutions (Johansson et al., 2012). Breidbach et al. (2013) found that people's motivation their actions play a more critical role in the ICT-enabled co-creation process than the technology itself. Technology acts as a facilitator rather than a moderator of change. The social collocations created and the decisions and actions taken by people play a pivotal role in this process.

In the field of urban policies, studies on dwellers' participation in the co-creation of solutions for cities show an increase in engagement due to both residents and the government side. As a result of collaboration, authorities become more open to the needs and expectations of residents, their decisions become more effective and efficient, while city dwellers become more satisfied and accepting of government actions regarding the functioning of the city (Agusti et al., 2014). Evidently, cocreation is no longer seen as a way of production, but as an expectation of multi-level sustainable development of the whole city, through strengthening urban functioning, urban planning, municipal governance, inhabitants development, and building association among individual residents and officials (Wamsler, 2016).

The connection of these fields has recently found more and more followers. Building ICT-based city solutions facilitated by a co-creation process is thus becoming more popular. Much of it is used by commercial organizations, focused on their own strategic goals, employed to drive the process. The research focuses on identifying factors relevant to the co-creation mechanism, considering the variety of actors, culture, commitment and attitudes (Akterujjaman et al., 2020).

### 2.2    Moral judgements

According to Rest et al. (1997) "moral judgment is a psychological construct that characterizes the process by which people determine that one course of action in a particular situation is morally right and another course of action is wrong". Moral judgements are a pivot part of our engagement with society. They play a substantial role in how we perceive ourselves, others and the situation where we find ourselves. They allow us to decide whether to engage in an activity or withdraw. Our moral system also largely determines the climate (in)action we take (Peeters et al., 2019) and how we judge the pro-environmental behavior of others. Research indicates that people consider the intentionality of action when assessing its environmental impact. An activity for which both the purpose and effect were pro-environment was rated as less significant than one for which only the result was pro-environment, while the reason was not environmental (Hoogendoorn et al., 2019).

Markowitz (2012) argues that when people view environmental issues through the lens of moral judgment, rather than simply a scientific or technical problem, their proenvironmental attitudes are stronger. These people have a greater willingness to engage in climate change mitigation. They also express a belief that actively responding to climate problems is their ethical obligation. Likewise, Wang (2017) found that peoples' moral attitudes correlate with subjective norms. Moreover, moral attitudes correlate with anticipated guilt. Individuals with collectively higher moral attitudes, higher levels of anticipated guilt, low empathy, and future orientation were more likely to show greater willingness to engage in global warming mitigation behaviors.

## 3      Methodology

### 3.1     Research goals

The study serves the effective implementation of the project aimed at creating an educational application based on ICT solutions (called Greencoin application), which is intended to be a tool for shaping and strengthening the pro-environmental behavior of city residents. The study methodology, regarding the consecutive phases of the project, follows the steps presented in Figure 1. The first phase is to define the main objective of the project: create a mechanism based on an interaction between government, private sector and citizens to bring sustainability to our cities. The proposed system will be implemented and tested in the Gdańsk and then can be transferred into any other city.

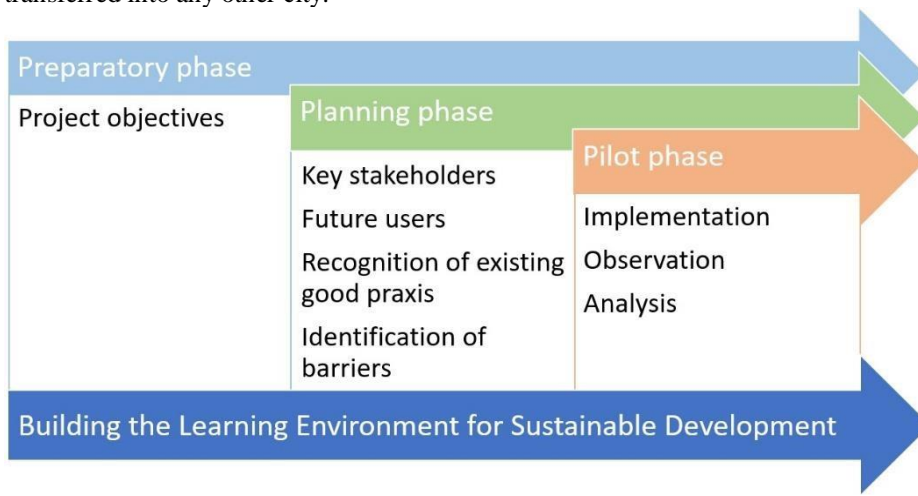

**Figure 1. The study framework. Author's own elaboration**

In order to achieve the project objective, it was planned to involve a wide range of stakeholders in the participatory process of active creation of solutions initiating pro environmental behavior among city dwellers. Workshops for crucial stakeholders and future recipients of project activities will enable the establishment and maintenance of an active dialogue with them at every stage of the project flow. A direct relationship will facilitate the creation of a product that meets not only the needs of the environment but also the current needs of city dwellers.

The purpose of the current phase of the project, and thus of the presented study, is twofold:

(1)    to identify existing beliefs and moral judgments that may facilitate the implementation of ICT-based educational activities that shape pro-environmental attitudes and behaviors among urban residents;

(2)    to identify beliefs and moral judgments that may hinder the implementation of ICT-based educational activities that shape pro-environmental attitudes and behaviors among urban residents.

### 3.2    Participants

A co-creation methodology was used to achieve the objectives of the study. The second phase of the project employed a series of participatory workshops. The first one-day workshop was held stationary (Figure 2). Due to restrictions related to the state of the epidemic, subsequent workshops were held remotely. The cycle of workshops was conducted between October and November 2021. The stationary workshop took place in the city of Gdańsk. The next ones in remote form allowed the participation of residents of other Polish cities. The workshops were attended by local stakeholders active in different thematic areas and occupying different socioprofessional functions. About 50 participants took part in the entire cycle of workshops.

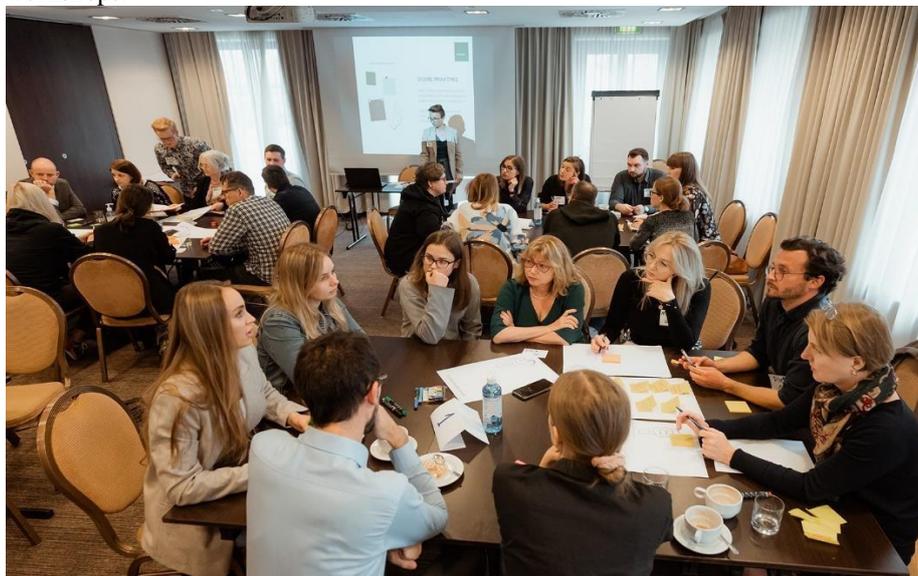

**Figure 2. Photo from the co-creation workshop held onsite**

### 3.3 Workshops

The workshop aimed to provide information on the issue of implementing an alternative currency as a solution to support the development of informal networks and services between local community residents. The workshop was held in five sessions, face-to-face for one hour each over one day, remote for 20 minutes each, over one afternoon. The main issues of the workshop included activities aimed at identifying difficulties and challenges faced by city dwellers, identifying wellfunctioning solutions in Poland and the world, reflecting on ways to solve current challenges of cities, in response to the determined challenges reflecting on the system of the functionality of the planned application and the set of rewards supporting it, analyzing the strengths and weaknesses of the potential mechanism for supporting pro-environmental behavior for city dwellers.

The workshop structure consisted of one introductory lecture, presented with the help of a power-point presentation, and other practical sessions, carried out in a stationary form in groups of several people, at tables, using sheets of paper, stickers and colorful stationery. Participants in the remote workshops used the equivalent tools offered by the Jambord application. The workshop was also attended by members of the project team. Four of them joined each group of workshop participants, one acting as moderator, one as a facilitator, the other two as observers. They observed, listened and noted down the participants' perceptions and behaviors accompanying the discussion of the workshop topics, including uttered beliefs, moral judgements and comments.

### 3.4 Data analysis

The data gathered included material collected by six observers (members of the project team) during the onsite and two observers during the remote workshops. The material consisted of notes taken directly during the workshops and after the workshop during the project team's closing discussion. Data analysis was conducted using a general inductive approach strategy (Thomas, 2006). According to the procedure adopted, coding was done based on the raw data. The categories were taken from the dominant and accented phrases uttered by the workshop participants. The advantage of this approach is that "although the findings are influenced by the evaluation objectives or questions outlined by the researcher, the findings arise directly from the analysis of the raw data, not from a priori expectations or models" (Thomas, 2006, p. 239). The codes selected reflect the actual content of the respondents' statements. The analysis carried out was aimed at developing the detailed beliefs, moral judgements reported directly by the workshop participants.

# 4 Results

## 4.1 Responsibility

The workshop participants' statements indicate a divergence in beliefs and moral judgements of various stakeholder groups. The category that emerged most often was the attribution of responsibility for caring about the quality of the urban environment. The example cited was waste management. Municipal officials undertake various policies and top-down remedial measures to improve the situation in the city, but these do not have the expected, noticeable effect. They take the form of regulation through appropriate legislation specifying waste collection and disposal fees. Despite the necessity of waste segregation into particular fractions, the city does not have a good way of verifying how this process works. It is particularly evident concerning housing estates, where there are no mechanisms to control whether and how residents separate waste. Officials limit their actions to establishing appropriate guidelines but do not take successful measures for their implementation, indicating that the responsibility for segregation lies with the residents themselves.

On the other hand, the residents feel responsible for proper waste segregation to a limited degree, shifting the responsibility onto companies collecting waste from their properties. Workshop participants pointed out that the city authorities do not carry out appropriate educational activities to raise the inhabitants' awareness of both the correct methods of waste segregation and the consequences of such actions for the environment but also the economic aspects of the city's operations. If the city managers were responsible for their waste management policy, implementing solutions facilitating its efficient implementation, the inhabitants would be more motivated to cooperate for the common good. Another aspect is the inadequate selective waste collection system in public spaces and companies/institutions. The users of these spaces do not feel responsible for looking for the right bin to dispose of a given fraction. They use those bins which are closest.

Other issue is the responsibility for costs. Despite the implementation of European Union directives, inhabitants are faced with the dilemma of who should bear the costs of producing excessive amounts of waste. Workshop participants noted that the trend of producing packaging that is disproportionately to the size or needs of the product, overuse of pre-packaging, or multiple packaging of the same product, serving de facto to manipulate consumer behavior, is growing. They blamed the authorities for the lack of an effective law forcing producers to use returnable packaging. On the other hand, the customer does not have much influence on how the product is packaged. Workshop participants expressed the belief that the producer should bear the costs of waste utilization, as he decides on the form of packaging produced, which soon becomes rubbish hard to recycle.

Urban transport is another example of the blurring of the responsibility for adopting pro-environmental behavior. Workshop participants representing city residents pointed out that the main reason for poor mobility habits is poor city management. The location of strategic services is mainly in the central districts. The granting of building permits for numerous new apartment blocks without the simultaneous construction of schools and kindergartens means that residents are forced to commute constantly to the city center - resulting in heavy traffic jams. The inflexible public transport fare system exacerbates the situation. The city authorities are unable to carry out an effective process of integrating urban transport services offered by dispersed providers. Overcrowded public transport discourages people from using public transport.

## 4.2    Sense of empowerment

Another category of beliefs that emerged from the workshops was that of empowerment. Representatives of city residents repeatedly expressed the belief that they have little influence on the quality of the environment. They perceived their activities as insignificant. They said they were not very motivated, for example, to save water in their flats, seeing that other neighbors in the block of flats did not do so. In addition, water consumption for shared purposes, the costs of which are distributed among the inhabitants of individual flats (washing of staircases, washing of the nearest neighborhood belonging to the block of flats, watering the greenery around the block of flats by the managers) generates high charges, concerning which individual savings are not satisfactory.

Home heating was another example. Residents of detached houses spoke about how difficult it was for them to give up wood-burning, especially in their fireplaces. They claim that they have good quality fireplaces certified by law and do not emit as much smoke as the industrial plants in the city. They do not use them constantly, often once or twice a week, especially in the early heating season. The reason for using fireplaces is for comfort purposes and aesthetics rather than for actual heating needs, which creates the impression of a negligible global impact on the environment. Despite the interest in environmental issues, this behavior was not seen as being unecological, especially as wood is still considered by the population as a renewable resource.

## 4.3    Local leadership

The next category that emerged was the attribution of blame for the lack of action to an insufficient number of community leaders. Workshop participants felt that they would like to do something to improve the quality of the environment in which they live, but they do not know what exactly to do and how to do it. According to them, their closest environment lacks initiators of social change. In particular, the fact that their employers are not such initiators was highlighted. If, for example, employers support their employees through a commuting subsidy scheme, they would be more likely to use city transport or cycle. Workshop participants also cited examples of shared commuting across borders. Several people use their means of transport to reach an agreed point, where they change to one car and travel together to their workplace in the city center.

Another example is the lack of positive change leaders among the managers of large corporations or workplaces, who do not try to be truly eco-friendly, often using sham measures. Positive solutions developed in the pandemic period are abandoned in the post-pandemic period. An example of this was the circulation of documents. At the time of the pandemic, most workplaces introduced an electronic workflow confirmed by an electronic signature. Employees started to use it in their day-to-day work, and it seemed that due to the numerous perceived advantages (saving time, paper, toner) this system would become permanent. However, with the return to stationary work many employers began to require paper documents again. According to the workshop participants, if the employer is not an initiator or an eco-leader, the employees will not implement environmentally friendly solutions in their daily work either. The potential of employers is untapped in this area.

### 4.4    Real eco-approach

The next category indicated by workshop participants was insufficient social acceptance of eco-behavior. Without positive experiences related to the benefits of displaying pro-environmental attitudes and behaviors, city dwellers have little motivation to demonstrate them in everyday life. Respondents clearly expressed the belief that there is a lack of consistency in the media approach to environmental problems. On the one hand, the media publish many films, documentaries, and social actions aimed at raising the ecological awareness of their audiences and encouraging them to implement positive changes to improve the quality of the environment. On the other hand, media flood their audiences with aggressive advertisements tending to increasingly consumerist lifestyles. What is more, these advertisements are often targeted at the youngest viewers.

Workshop participants were concerned about environmental issues but felt that their environmental awareness was limited and vulnerable to manipulation. They gave the example of a recent advertising campaign promoted in the media. A chain of shops encouraged people to buy water under the slogan "save wolves", declaring that they would donate a certain amount from the sale of each bottle of water to this cause. However, while helping nature in this way, they were also harming it because not only did they encourage people to buy bottled water instead of promoting drinking tap water, but they were also selling it in bottles made of plastic.

During group discussions, participants expressed the opinion that they would like to live a greener lifestyle, but it takes too much time. They also felt that the costs of being green are passed on to consumers, which does not motivate behavior change. Individuals represented different areas and professional positions, which was reflected in what they saw as the main environmental challenges in the city. Officials expected residents to be more involved in the actions they promoted or implemented, while residents expected the apparent removal of barriers limiting the implementation of green behaviors. Participants suggested that the benefits of implementing new solutions would be higher if we take care to co-create greater social acceptance of pro-environmental approach.

### 4.5    Eco-knowledge

The last category that emerged most frequently was adequate knowledge of environmental issues. It was perceived two fold. Firstly, there was a discussion on the difficulty of building de facto environmentally friendly solutions. Participants pointed that the basis for creating such solutions is in-depth knowledge of the course of individual processes. An application that could monitor mobility behavior was used as an example. Creating a system that controls whether a user earns points for actually riding a bicycle (and not, for example, a scooter or riding a bus that moves slowly because of a traffic jam) requires the implementation of solutions based on a network of Global Positioning System (GPS) transmitters, software, and devices that consume large amounts of energy. Expert knowledge is required to create a system whose operating costs based on excessive energy consumption, the transmission of large amounts of data, and network traffic consumption will not outweigh the profits from users' ecological behavior.

In-depth knowledge is also needed for the optimal inclusion of stakeholders in the created cooperation network. Participants discussed how to assess whether a potential project partner is environmentally friendly, while it may be a big polluter and have a negative environmental impact. City dwellers often have difficulty confirming what is green behavior. An example cited in this regard was electric cars. In the media space, various reports are provided on whether electric vehicles are really green because the electricity to power them can come from burning coal or gas. So, by choosing to buy an electric car and then powering it with electricity of unknown origin, are citizens contributing to environmental degradation?

Workshop participants also pointed out the threat of the growing phenomenon of greenwashing. Companies, driven by the desire to increase profits, deliberately mislead their customers by suggesting that the solutions dedicated to them or the products offered are ecological. In reality, customers are unable or cannot verify this. Negative experiences cause a lack of trust in subsequent campaigns or offers, and people do not believe that ecological products are really eco-friendly.

## 5       Discussion and conclusions

Creating a learning environment for sustainable development, of which the ICT-based planned application (called Greencoin) is an example, is a process that requires consideration of many factors that affect its subsequent effective functioning. The selection of active collocations and interactions between stakeholders and future users should consider not only the needs and expectations of particular groups of users but also their beliefs or moral judgments about environmental issues. The conducted workshops for stakeholders and future users of the application gave the representatives of these groups a voice in the co-creating process of environmentally friendly solutions, which have a chance to become technically feasible and socially acceptable at the same time. Taking into account the views and beliefs of city dwellers in connection with the experienced problems of everyday urban life facilitates the creation of a pilot solution that can be replicated. When implemented in other cities, it will shorten the process of reaching out to people so that further users can be involved more quickly and effectively to improve environmental quality.

The presented research was aimed to identify existing beliefs and moral judgements that can facilitate the implementation of ICT-based educational activities shaping pro-environmental attitudes and behaviors among city dwellers. Conducted workshops revealed the need to recruit local leaders for the implementation process of the application, who will enable to reach the city dwellers and encourage them to join the project activities. This endeavor, considered a socio-economic investment, should bring together, as confirmed by other studies, both leaders at the local (Homsy, 2018), neighborhood (Kretser & Chandler, 2020) and workplace levels (Afsar et al., 2018; Kumar et al., 2022). Workshop participants attributed local leaders as initiators of social change, providing direction and a course of action, motivating people to become active in improving the quality of the environment (Gruber et al., 2017).

At the other extreme of the approach focused on looking for factors that support the process is the approach that looks for barriers preventing the process from taking place. Removing the obstacles identified will allow the process to happen naturally. According to this approach, the second aim of this study was to identify beliefs and moral judgements which may hinder the implementation of ICT-based educational activities shaping pro-environmental attitudes and behaviors among city dwellers. One of the barriers identified during the workshop was the attribution of responsibility for implementing pro-environmental behavior to other people. Residents pointed to city managers as those who should initiate change, both by introducing rational regulations governing the multidimensional functioning of the city and by implementing policies and initiatives for the efficient and ecological functioning of the city. On the other hand, city authorities or officials declared that the solutions they are creating are appropriate, but the inhabitants lack the will to act and the sense of responsibility for the environment in which they live. Just as Eckersley (2016) argues, a complex model of social connections based on a system of structural inequalities leads to a blurring of the notion of co-responsibility for climate change and even to the creation of a belief of collective irresponsibility among representatives of particular groups of social actors.

The findings have raised the question of how to strengthen the sense of responsibility for the environment among urban residents. A co-created Greencoin application could reach people who do not consider environmental care as their daily responsibility or who expect particular actions for environment from others (Cleveland, 2020). The theme of meager sense of empowerment also emerged concerning this matter. Workshop participants impugned the relevance of taking individual action due to its low effectiveness, especially against the background of global anti-environmental endeavors. Questions arose: Is it even worth it to act individually? Do our efforts make sense? Will anyone appreciate our involvement? It is not uncommon that people decide to take pro-environmental actions, but when they see the low effectiveness of those behaviors, it decreases their motivation to act for the benefit of climate, so they abandon the undertaken direction of activity (Abrahamse, 2019).

The results obtained indicate that the co-created functionalities of ICT applications for sustainable development should consider the issue of knowledge transfer, which is the basis for building a pro-environmental approach in society. We live in a world where the problem is not a lack of information but an excess of it. We are dealing with a twofold situation here. Knowledge is developing so fast that it is impossible to be an expert in many areas. People's knowledge about climate and the environment is often too superficial, making them susceptible to commercial manipulation or the negative influence of eco-sceptics (Lukinović & Jovanović, 2019). The lack of indepth knowledge also causes difficulty in assessing whether the taken actions and the resulting consequences are, in fact, environmentally friendly.

The currently developed Greencoin application, like other similar solutions, should consider the results obtained, representing the beliefs, opinions and judgments held by representatives of the local community. If they are not taken into consideration at the solution development stage, then the whole effort of the project team could turn out to be useless, as the product created will not be adapted to the needs and expectations of the future user. Our responsibility is to develop technology that responds to the challenges of this world and is a tool that does not drive change, but enables the change.

The limitation of the presented study is twofold. Firstly, the study was conducted in Poland as an initial stage of an ongoing project. The results are currently of local nature. Extension of the research scope is planned after the pilot phase. Secondly, the research material was gathered by different observers participating in three independent, parallel groups of workshop participants and therefore is inevitably subject to perceptual bias. Therefore, a follow-up study is envisaged and will be based on a series of in-depth focus interviews. Their analysis will be presented in a future article.

## Acknowledgments

This research is supported by €1.9 million in funding received from Iceland, Liechtenstein and Norway under the EEA Funds, grant agreement NOR/IdeaLab/GC/0003/2020-00.